# Influence of Memory Hierarchies on Predictability for Time Constrained Embedded Software*

Lars Wehmeyer, Peter Marwedel
Embedded Systems Group, CS Dept., University of Dortmund, Germany
{Lars.Wehmeyer, Peter.Marwedel}@udo.edu

## Abstract

*Safety-critical embedded systems having to meet real-time constraints are expected to be highly predictable in order to guarantee at design time that certain timing deadlines will always be met. This requirement usually prevents designers from utilizing caches due to their highly dynamic, thus hardly predictable behavior. The integration of scratchpad memories represents an alternative approach which allows the system to benefit from a performance gain comparable to that of caches while at the same time maintaining predictability. In this work, we compare the impact of scratchpad memories and caches on worst case execution time (WCET) analysis results. We show that caches, despite requiring complex techniques, can have a negative impact on the predicted WCET, while the estimated WCET for scratchpad memories scales with the achieved performance gain at no extra analysis cost.*

## 1 Introduction

The growing gap between increasing processor speeds and the slower main memory, also known as the "memory wall" [1], has become a bottleneck for computer designers. In traditional computer systems like desktop PCs, caches are usually introduced in order to hide the high latencies of main memory accesses. Especially for real-time embedded systems, predictability is an important issue: even during the design phase of such a system it must be guaranteed that certain deadlines will always be met. The use of caches, however, tends to improve only the average-case performance, not necessarily the worst-case execution time (WCET). Estimating a cache's contribution to WCET requires complex analysis techniques to model its dynamic behavior which is hard to predict in a safe, yet not over-pessimistic way.

An alternative approach that has been investigated with respect to performance and energy consumption is the use of so-called scratchpad memories, also known as "tightly coupled memories" (TCM). They are small on-chip memories mapped into the processor's address space. Due to their small size, scratchpad memories are extremely fast and require very little energy per access. They are more efficient than caches since the hardware logic required to control a cache is not required for a scratchpad. The organization and utilization of the scratchpad is rather left to the programmer or to the compiler. However, a comprehensive methodology for the efficient utilization of scratchpad memories is surprisingly still missing in industry.

Previous work proposed mapping the hot-spots of an application to the scratchpad memory in order to gain performance and to save energy. Apart from improvements concerning these optimization goals, the mentioned methods also have a beneficial effect on worst case execution time. In contrast to a cache, all decisions concerning the layout of memory objects during execution of the application are fixed at compile time, making all memory accesses inherently predictable. In this paper, we assume different sizes of scratchpad memories and unified caches. For each configuration, we determine the performance of several benchmarks by simulation using a typical input data set. Additionally, we perform WCET analysis on the applications, once for scratchpads and once for caches. The results obtained for scratchpad memories and caches are then compared.

The rest of this paper is structured as follows: the next section considers some of the previous work on scratchpad allocation algorithms as well as WCET analysis, which form the basis of this work. In section 3, we describe the workflow used to generate application executables for use with scratchpad memories and caches and show how the WCET analysis was performed. Results are presented in section **4**. A summary and possible future work conclude the paper.

## 2 Related Work

Today's markets expect computer systems to show a steady increase in computing power. Since the limitations concerning miniaturization and ever faster gigahertz processors are starting to show, computer architects are forced to include performance enhancing features so as to meet the customers' demand for high performance. Examples for such features are the use of pipelines or speculative execution using branch prediction units. The growing speed gap between processors and the slower memory is the main reason for the widespread integration of caches.

All the above techniques help increase the average case performance of a system. In embedded systems, however, it is often necessary to be able to guarantee that timing deadlines will never be violated. The required worst case execution time analysis techniques become increasingly difficult when many of the above-mentioned architectural features are present in the

---

processor. Their highly dynamic behavior makes it hard or even impossible to effectively predict the worst case timing at design time [2]. A general overview over available analysis techniques for the architectural features mentioned above can be found in [3]. Of particuar interest is the work by Li et al. [4], who consider the presence of a direct mapped instruction cache in the WCET analysis of embedded systems. A cache conflict graph is used to approximate the behavior of the cache and to determine the total number of hits and misses. A follow-up paper [5] extends the work to also cover set associative instruction caches as well as data and unified caches. One solution to the problem of data caches presented in [6] is the introduction of predictable data structures, which should be used by the programmer for timing critical code. Tan et al. [7] extended the consideration of caches to also cover the case of multi tasking systems. In this case, the preemption of tasks can lead to additional cache miss overhead which has to be considered and evaluated, further complicating cache analysis.

A concept for separating program path analysis and microarchitectural analysis into two steps in order to reduce the complexity of WCET analysis is presented in [8]. Results are reported to be comparable to combined analysis techniques. This approach is also used in aiT [9], a commercial WCET analysis tool that is available for several processor and cache architectures. aiT is actively used in industry, e.g. by Airbus France in order to determine upper bounds for the execution times of critical avionics software. As input, the tool takes an executable for a specific platform along with user supplied annotation data concerning e.g. loop bounds and access addresses as well as architectural information concerning the memory layout. It then generates a safe upper bound for the expected WCET. Using aiT, the elaborate (if at all feasible) task of finding input sets for which a simulation run yields the maximum execution time is no longer required. The commercially available version of aiT for ARM7 is currently not equipped with a cache analysis. AbsInt GmbH provided us with a simple experimental cache analysis for the ARM7 cache that uses only a subset of the analysis techniques [10] available with commercial versions of aiT. One of the difficulties with caches integrated into ARM processor cores is the fact that they use a random replacement policy, making precise estimates for cache behavior difficult. For caches that use an LRU replacement, WCET analysis can yield tighter bounds.

In contrast to a cache, no extra analysis module is required in order to investigate the effect of a scratchpad memory on WCET, since it is simply introduced as a new, distinct memory region.

In general, scratchpad memories are an effective replacement for caches since they can help bring down energy consumption and at the same time offer performance benefits comparable to those of caches [11]. For this reason, scratchpad memories are becoming more popular and are widely available e.g. in the ARM9 processor series under the name Tightly Coupled Memory (TCM). The one drawback of scratchpad memories is the fact that they need to be actively exploited by the programmer or the compiler. Since the scratchpad does not have any logic to dynamically control its contents at runtime, memory objects have to be allocated to the scratchpad by the compiler. The handling of memory allocation during the compilation process is advantageous since the compiler has detailed knowledge about execution and access frequencies. This information is used to distribute memory objects among the available different memories in an optimal way instead of using dynamic ad-hoc decisions as in a cache.

Memory allocation can be performed either in a static or in a dynamic way. In the former approach, the scratchpad is preloaded with a set of memory objects which stay on the scratchpad throughout the application's execution time. This static approach was first used in [12] to allocate data objects like arrays to a scratchpad memory. In [13], both instructions and data are allocated to the scratchpad in order to save energy by exploiting the scratchpad memory's low power dissipation.

The dynamic approach allows memory objects to be copied to the scratchpad at runtime. Dynamic allocation techniques presented in [14, 15] consider data and instructions, respectively. These approaches are of particular benefit when large programs with several hotspots and changing working sets are used. A new paper [16] uses a technique based on register allocation for CISC architectures to allocate both instructions and data to the scratchpad in a dynamic way. Considerable savings of up to 38% concerning energy consumption compared to a static approach are reported. Note that the objects and the points in time at which they are copied to the scratchpad are all fixed at compile time. Thus, both static and dynamic scratchpad usage are under full control of the compiler or the programmer, making the methods inherently predictable. Concerning WCET, this means that using a scratchpad and a compiler-based algorithm to utilize it, no additional analysis is required for WCET analysis. Despite the additional overhead at compile time for application analysis and object allocation to the different memories, using a scratchpad offers an advantage over caches which generally show a dynamic behavior that is hard to predict at compile time and require complex analysis techniques in order to determine a tight upper bound for WCET.

In this paper, we will use the static allocation technique presented in [13] to allocate functions and global data to the scratchpad in an energy-optimal way using the energy model for the ARM7TDMI processor described in [17]. The energy consumption of caches and scratchpad memories is modelled according to [11]. The problem of finding an optimal mapping of memory objects to the scratchpad and main memory is solved by formulating it as a variant of the knapsack problem and using a commercial ILP solver [18]. This is repeated for different scratchpad memory capacities. The resulting executables are simulated with ARM's instruction set simulator ARMulator [19] using a typical input data set leading to an average case runtime.

To determine the effect of using scratchpad memories on WCET, the executables using different sizes of scratchpads are also analyzed using aiT. Information about the used memory architecture, including main memory and scratchpad memory timing and address information, has to be provided to the tool. Apart from this annotation, no further information or analysis is required compared to a system that only uses main memory.

To compare scratchpads with caches, an executable generated without using the scratchpad optimization in the compiler is simulated using unified, direct mapped caches of different sizes. These parameters are configurable in ARMulator. WCET analysis is performed using the available cache analysis tool for the ARM7 processor integrated into aiT. The determined WCET using a cache is then compared to the scratchpad case.

## 3 Workflow

The workflow used to compare the impact of scratchpads and caches on WCET analysis results is shown in Figure 1. The benchmark programs, written in the C programming language, were compiled into executables assuming either a cache or a scratchpad in the target system. Our encc compiler generates instructions in the 16 bit THUMB mode of the ARM7 processor. Because of the higher code density, this instruction set is recommended for energy- and size-constrained systems [20].

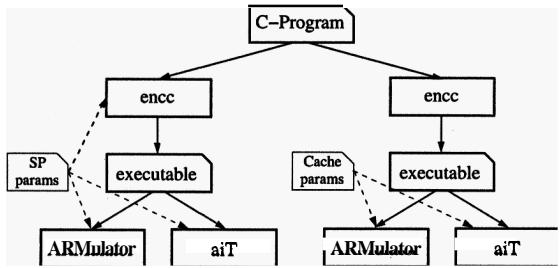

Figure 1: Workflow

The left branch of Figure 1 shows the scratchpad setup. To generate simulation results for a scratchpad based system, the compiler takes as input the size and the access costs of the scratchpad to be used. This information is used to solve the corresponding knapsack problem. It is formulated in ILP notation using a benefit function which associates each memory object (function or global data element) with a certain energy gain if this object is statically allocated to the scratchpad memory instead of main memory. This benefit is maximized in the objective function under the constraint that the capacity of the scratchpad is not exceeded. For details of the used algorithm, please refer to [13], which also describes how to treat basic blocks and multi basic blocks, an extension not used for the results in this work.

The generated executable then contains address information for all memory objects, which means their static location in main memory or scratchpad is known. The executable is simulated using ARMulator, which in turn receives information about the size and the address range of the used scratchpad memory. ARMulator can thus determine the number of cycles required to execute the benchmark using a typical example input data set, taking into account the reduced access latencies of the scratchpad memory compared to main memory. The result of this step is a simulated average case execution time. This code generation and simulation was repeated in our workflow for scratchpad sizes from 64 bytes to 8k.

The influence of using a scratchpad on WCET was determined using aiT. Even though no additional software module is required in order to investigate this effect, the use of a scratchpad requires an annotation in the aiT configuration files. aiT supports the specification of memory regions with different attributes. As shown in table 1, the scratchpad region always requires one cycle per access. Annotating the timing of the scratchpad region is actually the only additional effort required in order to support scratchpad memories in the WCET analysis.

| Access Width | Main Memory | Scratchpad |
|---|---|---|
| Byte (8 Bit) | 2 | 1 |
| Halfword (16 Bit) | 2 | 1 |
| Word (32 Bit) | 4 | 1 |

Table 1: Cycles per memory access (access + waitstates)

In our model, which is based on the AT91EB01 evaluation board by ATMEL Corp., the access times to main memory depend on the width of the access. Since aiT does not currently allow the definition of access times depending on the bit width, the waitstates for the different regions in main memory have to be annotated in order to obtain valid results concerning WCET. Since our compiler generates 16 bit THUMB mode instructions, but integer data elements occupy 32 bit, the different access times for instruction and data fetches shown in table 1 have to be accounted for. An instruction fetch from main memory causes one cycle for the actual access and one additional waitstate. Accessing a 32 bit data element from main memory requires four cycles, since three additional waitstates occur. This is reflected in the annotation file shown in Figure 2 for one benchmark and one particular scratchpad configuration: The first memory area represents the scratchpad memory: Each access takes 1 cycle with no additional waitstates, independent of the bitwidth of the access. The "1:1" indicates that the memory runs at the same clock speed as the processor. For main memory, we need to differentiate between 16 and 32 bit accesses. The second region contains 16 bit instructions, requiring two cycles per access. The next region represents a so-called literal pool, 32 bit data elements within the instruction region used to load large constants into a register. Accessing a 32 bit value from a literal pool requires four cycles.

```
# Scratchpad
MEMORY_AREA: 0x400eb0 0x400fbb 1:1 1 READSONLY CODE&DATA
# Main memory regions
# Instructions
MEMORY_AREA: 0x400fbc 0x401063 1:1 2 READ-ONLY CODE-ONLY
# Literal Pool
MEMORY-AREA: 0x401064 0x40106f 1:1 4 READ-ONLY DATA-ONLY
# integer data
MEMORY_AREA: 0x401070 0x402083 1:1 4 READ&WRITE DATA-ONLY
# array of short
MEMORY-AREA: 0x402084 0x402092 1:1 2 READ&WRITE DATA-ONLY
```

Figure 2: Example annotation for aiT using scratchpad and main memory

The remaining memory regions represent the data region of the executable. Arrays of 32 bit values require 4 cycles per access, whereas an array of variables of type short (**16** bit) only takes **2** cycles. The regions are annotated accordingly.

The specification of the memory areas requires some annotation overhead, but it should be noted that this is always necessary if memories with different access times are used. The scratchpad memory with its uniform access times actually simplifies the annotation process since scratchpad regions are not divided depending on the bit width of the memory objects. Most of the regions and addresses can be determined automatically from address information provided by the linker.

One more requirement for a scratchpad memory is the annotation of function calls that jump from main memory to the scratchpad or vice versa, since the relative branch offsets within the executable do not reflect the actual execution time addresses after the program has been loaded into memory.

We will now consider the workflow employed when a cache is used in the memory hierarchy (right branch of Figure **1**). Since a cache is in general transparent to software, it does not need to be considered during code generation. Therefore, generating one executable for use with all cache sizes is sufficient. Of course, there are cache-based optimizations that can help prevent cache conflicts through techniques like array partitioning or loop tiling [21], but these are not considered in our approach since it is doubtful that the benefits will also reflect in improved WCET estimates.

The executables generated by the compiler are simulated using ARMulator. The instruction set simulator requires information about the cache size and organization in order to be able to determine the number of cycles required for executing the benchmarks for the different cache capacities from **64** bytes to 8k. For our experiments, we assumed a simple direct mapped unified cache architecture found in ARM processors. Each cache line holds four **32** bit words.

To determine the WCET using caches, the executables are also analyzed using the WCET analysis tool aiT. The cache analysis feature for the ARM7 is used to estimate the WCET of the benchmarks using different cache sizes. Despite the fact that the used cache analysis only includes a MUST-analysis and no persistence considerations, we would like to stress that for a scratchpad, no additional analysis technique is required at all.

Like for the scratchpad case, information about the timing of cache hits and misses has to be annotated in the aiT configuration files. The differentiation between **16** an **32** bit accesses required above for main memory accesses is not necessary in this case, since the cache always performs **32** bit accesses to fill an entire cache line on a miss. Assuming a cache line length of four, the loading of an entire cache line requires four **32** bit accesses to the main memory. According to table **1**, this means **12** additional waitstates, assuming the used memory does not support burst transfers. A cache hit only requires one cycle to retrieve the accessed word from the cache. This timing information is annotated in aiT's configuration file.

In addition to the memory region annotations mentioned above, the user also needs to specify the bounds of loops that aiT did not detect automatically, as well as the range of possible addresses for those array accesses that could not be determined by aiT. This can happen within our framework because starting addresses of arrays may be allocated to a different memory region (e.g. to scratchpad), which is not directly reflected in the executable analyzed by aiT. Array access annotation is of particular interest for the cache analysis, since it is in general not decidable which element of the array is actually being accessed at what time during execution of a benchmark. The generation of all of the annotations mentioned above is automated using information from the simulator and from the linker.

Once all annotations have been performed, aiT can analyse the WCET of the application. The results of simulation and WCET analysis for scratchpad memories and caches are presented in the following section.

## 4 Results

The benchmarks used to explore the impact of scratchpads and caches on WCET are given in table **2**. They comprise two speech encoding and decoding algorithms from the mediabench benchmark suite [22] and a mix of sorting algorithms commonly found in many algorithms.

| Name | Description |
|---|---|
| G.721 | Speech encoding and decoding, reference implementation of the CCITT |
| Multi_Sort | Combination of sorting algorithms |
| adpcm | Speech encoding and decoding using Adautive Diff. Pulse Code Modulation |

Table 2 Benchmarks

For all results presented in this section, there is always a certain difference between the WCET estimated by aiT and the number of cycles determined using simulation. Please note that this overestimation is due to the comparison of an average case simulation using typical input data to the longest possible execution time. This approach was chosen since it is generally infeasible to determine a worst case input data set for an arbitrary application. Using a simple sorting algorithm with a known worst case input data set, the results obtained by simulation on one hand and by WCET on the other only differed by **0.2%**, highlighting the high precision of the used WCET analysis tool.

First of all, the benchmark programs were compiled and simulated using ARMulator with a setup of varying scratchpad sizes. As expected, the simulated execution time decreases when the scratchpad capacity is increased. So does the estimated WCET determined by aiT: as can be seen for the **G.721** benchmark in Figure 3a), WCET decreases at the same rate as the actual cycles determined by simulation do. The step function appearance of the scratchpad simulation cycle count is due to the consideration of only functions and global variables, as described above.

The next step comprises simulation and WCET analysis of a system employing a cache in the memory hierarchy – cf. Figure 3b). Simulation times using a cache are quite similar to the scratchpad values for the **G.721** benchmark. For a very small cache, the execution times go up due to the high num-ber of conflict misses. After that, times decrease at roughly the same rate as they do for a scratchpad. The estimated WCET, however, shows a very different behavior: it stays at a very high level for all cache sizes instead of scaling with the average case performance improvements. Despite the fact that aiT for ARM7 has only been equipped with a subset of the cache analysis techniques available with commercial versions of aiT for other processors, it should be clear that the much better results concerning WCET analysis when using a scratchpad were achieved with an even simpler analysis technique not requiring any additional cache analyses.

4Proceedings of the Design, Automation and Test in Europe Conference and Exhibition (DATE'05)
1530-1591/05 $ 20.00 IEEE

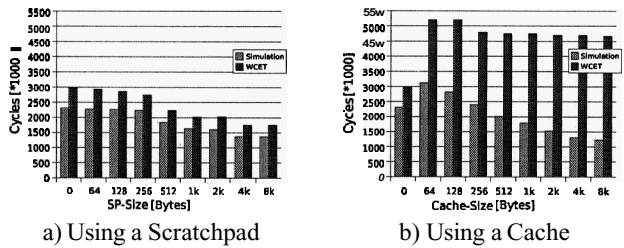

a) Using a Scratchpad    b) Using a Cache

Figure 3: Results for G.721 benchmark

Figure 4 shows the ratio of the WCET estimation to the simulated number of cycles for different scratchpad and cache sizes for the G.721 benchmark. The simulated number of cycles was normalized to the value 1. The main observation in this figure is that the difference between average case simulation and WCET analysis results remains constant for all scratchpad memory sizes, meaning the added performance obtained by including a scratchpad memory in a system translates directly to an improved estimated WCET. Using a cache, on the other hand, can lead to a strong WCET overestimation in particular for large cache sizes. Since the cache's behavior is hard to predict, it is difficult to provide a sufficiently tight upper bound for the WCET.

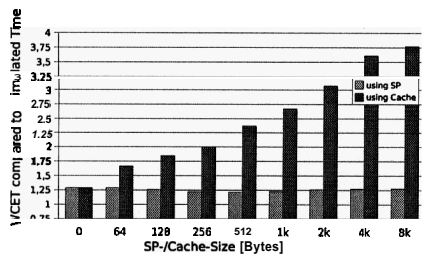

Figure 4: G721: Ratio of WCET and Simulated Cycles for cache and scratchpad based systems

A similar picture can be observed for the MultiSort benchmark. Using a scratchpad memory, the WCET is about 3 times as high as the simulation time for the given input data set. Note again that this overhead of WCET over simulation stems from the fact that typical, not worst case input values were used to generate simulation results. For the scratchpad approach, the ratio of WCET to simulation cycles stays more or less constant over the considered range from 64 bytes to 8k of scratchpad, whereas for the cache, the difference between WCET and simulation cycles increases strongly with the cache size. This is due to the ever larger uncertainty concerning cache misses when the cache size increases.

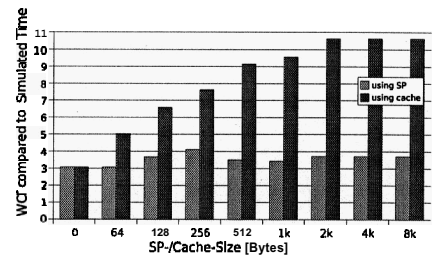

Figure 5: MultiSort: Ratio of WCET and Simulated Cycles for cache and scratchpad based systems

Figure 6 for the ADPCM benchmark shows a clear performance benefit of the scratchpad compared to a cache, in particular for small sizes. For a cache that is too small, a lot of cache misses occur for this program, leading to a severe performance degradation.

It can further be observed that the overall deviation of WCET and simulated cycles is always very **low** for this benchmark. This may either be due to the fact that the chosen input set is close to a worst case set, or that the program is not very control flow intensive and thus consists mainly of the critical path.

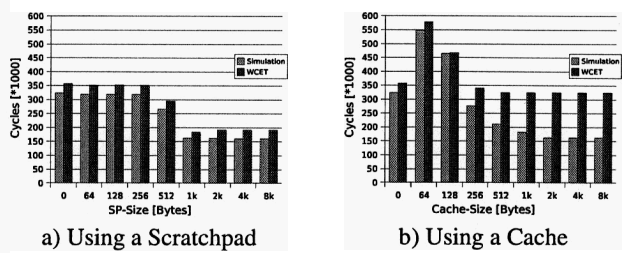

a) Using a Scratchpad    b) Using a Cache

Figure 6: Results for ADPCM benchmark

Despite the fact that the ratio of simulated average case time and WCET is indeed better for a cache of 128 bytes than for a scratchpad, it can be seen that both the performance and the WCET estimate for a scratchpad are better in absolute numbers. For larger cache and scratchpad sizes, the behavior seen for the other benchmarks also becomes apparent for ADPCM: the uncertainty concerning cache behavior prevents the WCET analysis results to be as close to the average simulation performance as they are for a scratchpad.

The results clearly show that using a scratchpad in a real-time embedded system is advantageous. Without further analysis effort, the WCET estimate reflects the actual average-case performance gain achieved by utilizing a scratchpad. The only necessary modification in the aiT tool is to specify the memory regions and their access latencies.

## 5 Summary and Future Work

In this work, we compare the effect of using caches or scratchpad memories on WCET analysis results for time constrained embedded systems. We use a known scratchpad allocation algorithm to place functions and data onto the scratchpad memory and annotate the configuration files of the used WCET analysis tool accordingly. We show that for caches, the difference between simulated average case execution time and WCET can grow larger for increasing cache sizes. The magnitude of this difference depends on the cache architecture and the analysis techniques. Using scratchpad memories leads to a decrease of the estimated WCET for growing scratchpad sizes, with a near constant ratio between measured average case simulation time and the WCET analysis results throughout the range of considered scratchpad sizes. The benefit obtained by using a scratchpad memory directly translates to a reduced WCET estimate. Scratchpad memories should thus be considered as a feasible and worthwhile option during the design of time constrained systems, since their application improves not only the average case performance, but also helps reduce the predicted WCET.

This work is a first step of comparing scratchpad memories and caches in real time embedded systems. In the future, we will consider other cache configurations (e.g. instruction caches instead of unified caches as well as set associative caches) to investigate their effect on WCET, and again compare the results to using scratchpad memories.

We expect that using the full scale of aiT's cache analysis techniques as described in [10] would probably lead to improved cache results with respect to WCET. However, despite the complexity of cache analysis tools available today, it is doubtful that the results achieved by using an inherently predictable scratchpad can be reached.

We will use the full featured allocation technique [13] also considering basic blocks for allocation onto the scratchpad memory instead of just complete functions. This is expected to further improve results for the scratchpad case due to the finer allocation granularity. The dynamic memory allocation of instructions and data presented in [16] will also be investigated under the aspect of worst case execution time.

Finally, the allocation technique will be extended to not optimize the allocation of objects to the scratchpad memory using an energy cost function, but rather to consider placing those objects onto the faster onchip memory that lie on the critical path of the application. This is expected to lead to even better WCET estimates.

## 6 Acknowledgement

The authors would like to thank "AbsInt" Angewandte Informatik GmbH for their support concerning WCET analysis using the aiT framework.